\documentclass[lettersize,journal]{IEEEtran}
\usepackage[utf8]{inputenc}
\usepackage{graphicx}
\usepackage{etoolbox}
\usepackage{setspace}
\usepackage{float}
\usepackage{amsmath}
\usepackage{svrsymbols}
\usepackage{titletoc}
\usepackage[titles]{tocloft}
\usepackage{algorithmic}
\usepackage{color,soul}
\usepackage{bibentry}
\usepackage{cite}
\usepackage{algorithmic}
\usepackage{algorithm}
\usepackage{amsmath,amsfonts,amssymb}
\usepackage{pgfplotstable}
\usepackage{booktabs}
\usepackage{array}
\usepackage{colortbl}
\usepackage{subfig}
\usepackage{multirow}

\newcommand*{\thead}[1]{\multicolumn{1}{c}{\bfseries #1}}

\begin{document}

\title{Detection of Malicious DNS-over-HTTPS Traffic: An Anomaly Detection Approach using Autoencoders}

\author{
\IEEEauthorblockN{Sergio A. Salinas Monroy, Aman Kumar Gupta, and Garrett Wahlstedt}
\thanks{This work has been submitted to the IEEE for possible publication. Copyright may be transferred without notice, after which this version may no longer be accessible}
\\
\IEEEauthorblockA{School of Computing\\
Wichita State University\\
}
}



\maketitle

\begin{abstract}
To maintain the privacy of users' web browsing history, popular browsers encrypt their DNS
traffic using the  DNS-over-HTTPS (DoH) protocol.  Unfortunately, encrypting DNS packets
prevents many existing intrusion detection systems from using plaintext domain names
to detect malicious traffic. In this paper, we design an autoencoder
that is capable of detecting malicious DNS traffic by only observing the encrypted DoH traffic.
Compared to previous works, the proposed autoencoder looks for anomalies in DoH traffic, and
thus can detect malicious traffic that has not been previously observed, i.e., zero-day
attacks.  
We run extensive experiments to evaluate the performance of our proposed autoencoder
and compare it to that of other anomaly detection algorithms, namely, local outlier factor,
one-class support vector machine, isolation forest, and variational autoencoders.  We find
that our proposed autoencoder achieves the highest detection performance, with a median F-1
score of 99\% over several types of malicious traffic. 
\end{abstract}
\section{Introduction}\label{sec:introduction}
The domain name system (DNS) is the Internet's core function that allows users to find web
servers using domain names instead of IP addresses. 
Although DNS offers many advantages, it allows eavesdroppers to observe
users' browsing history in plaintext. This is a critical issue because browsing history
contains sensitive information that users would prefer to keep private \cite{leith2021web}. 

To protect the privacy of users, most major browsers, including Mozilla's
Firefox and Google's Chrome, \cite{firefoxDoH,chromeDoH}, have adopted DNS-over-HTTPS (DoH). 
DoH encrypts the DNS traffic by first establishing a TLS connection between the user's web
browser and one of multiple available DoH public servers \cite{adguardList}. 
The browser then exchanges DNS packets encoded as HTTP requests with the DoH server.
Since the encrypted packets are only accessible to the user and the DoH server, DoH prevents
network administrators, ISPs, as well as malicious actors on the Internet from accessing the
user's DNS traffic, thus protecting their privacy. 

However, by encrypting the DNS packet payload, DoH interferes with the ability of existing
intrusion detection techniques that rely on observing domain names
\cite{bilge2011exposure,shi2018malicious,antonakakis2012throw,chin2018machine,
ahmed2019monitoring,zago2020scalable,woodbridge2016predicting,mac2017dga,spooren2019detection,
tran2018lstm,drichel2020analyzing,sidi2020helix,liu2019ccga}.
For example, intrusion detection systems (IDS) implemented on devices from Palo Alto Networks,
Cisco, and Trellix inspect the requested domains in plaintext to determine if they are
likely to have been generated by a domain generation algorithm \cite{paloalto,cisco,trellix}. 
In addition, since DoH traffic is now part of a network's typical traffic,
malicious DoH traffic generated by malware can appear normal, further complicating its
detection. Thus, it is important to develop privacy-preserving intrusion detection systems that
can detect malicious DoH traffic without the need to access plaintext information in the DNS
packets. 

There have been some efforts to classify malicious and normal DoH traffic using machine
learning classification approaches
\cite{montazerishatoori2020detection,singh2020detecting,kwan2021exploring,zhan2022detecting,alenezi2021classifying,Zebin2022AnEA,AlHaija2023ALD,MoureGarrido2022DetectingMU,Wang2022FFMRAD}.
However, since these works require both malicious and normal DoH traffic samples to be
available during their training phase, they are unable to detect previously-unseen malicious
DoH traffic, i.e., zero-day
attacks. Furthermore, they only focus on detecting DNS tunneling traffic over DoH, ignoring other
types of DNS abuse such as domain generation algorithms (DGA) that are used by bots to find
their command and control servers and can be used as an earlier indication of compromised devices. 


In this paper, we propose an autoencoder that can effectively detect zero-day malicious DoH
traffic. The autoencoder is a type of deep learning model that can be trained to take normal
DoH traffic as input and accurately recreate it as its output. When presented with malicious
traffic, the autoencoder will recreate it with a large error. The difference in the
autoencoder's recreation error allows us to identify malicious DoH traffic, even if it has not
been observed during training. 
The autoencoder preserves the privacy of the users' browsing history since it only
uses statistical information about the network flows and operates without accessing the
plaintext DNS data in the DoH packets. 

To evaluate the detection performance of the proposed autoencoder, we first 
collect realistic normal and malicious DoH traffic. Our normal traffic dataset is generated by
automating a browser to visit a list of common websites while using different public DoH
servers. The malicious traffic dataset contains both DGA and tunneling traffic. The DGA traffic
was obtained by implementing several DGA algorithms and then automating their DoH requests. To
simulate a DGA that attempts to avoid detection, we generated automated DoH requests with random waiting
times between them. The DoH requests where sent both over independent and shared TLS connections.
These datasets are available through our Kaggle repository \cite{salinas2023dataset}.
We obtain DoH tunneling data from the datasets in MontazerShatoori et al.
\cite{montazerishatoori2020detection}, which include traffic generated by multiple tunneling protocols. 

We train multiple autoencoder architectures with the normal DoH traffic dataset
and identify the one that best recreates it.  The best autoencoder architecture 
achieves up to a median F-1 score of 99\% in identifying malicious DoH traffic. Compared to other
anomaly detection algorithms (i.e., isolation forests, local outlier factor, one-class
support vector machine, and variational autoencoders), the proposed
autoencoder achieves a higher median F1-score, accuracy, area under the curve (AUC), precision
and recall across all types of malicious and benign traffic.

We summarize our contributions as follows. 
\begin{itemize}

  \item We design a privacy-preserving anomaly detection autoencoder to detect malicious DoH
  traffic. To the best of our knowledge, this is the first work to
  investigate anomaly detection-based methods to identify malicious DoH traffic, and thus the first
  one that is designed to detect zero-day attacks.  

 

 \item We create a new DoH traffic dataset that includes both benign and DGA-generated
       malicious traffic. The dataset is publicly available \cite{salinas2023dataset}. 
 
 \item We conduct extensive evaluations of our proposed autoencoder and compare its results to
 that of existing anomaly-detection methods that could potentially be used to
 detect malicious DoH traffic, including local outlier factor,
 isolation forest, one-class support vector machine, and variational autoencoders with various
 hyperparameter configurations. 

\item Our results show that our autoencoder can detect malicious DoH traffic with up to 
 a median F1-score of 99\%, which is higher than that of the other evaluated anomaly detection models. 
 


\end{itemize}

The rest of the paper is organized as follows. Section \ref{sec:relatedWorks} describes the
differences between related works and this paper. Section \ref{sec:background} explains the
differences between DNS and DoH. In Section \ref{sec:threatModel}, we describe the assumptions about
the DGA and tunneling attacks considered in this paper. Section \ref{sec:autoencoder} presents our
autoencoder architecture and anomaly detection approach. Section \ref{sec:dataset} describes the
data used to train and test the autoencoder. Section \ref{sec:results} presents the results of our
experiments. We conclude the paper in Section \ref{sec:conclusions}. 

\section{Related Works}\label{sec:relatedWorks}
Detecting malicious DNS traffic has been studied in several works. However, there are only a few
works that investigate detection of malicious DoH traffic, and, to the best of our knowledge, none
that use anomaly detection autoencoders to address this problem. In this section, we provide an
overview of existing works on malicious traffic detection methods, and highlight their
differences with respect to this paper.

\subsection{Detection of malicious DNS traffic}
There is a rich literature on malicious DNS traffic detection using the domain names that
appear in DNS packets. We present a few of the most relevant ones. 
Antonakakis et al.
\cite{antonakakis2012throw} design an approach that learns the textual characteristics of the
DGA-generated domain names obtained from non-existent domain DNS packets (NXDOMAIN).  
Schuppen et al. \cite{schuppen2018fanci} develop FANCI, which is one of the best performing
algorithms in the literature. FANCI uses the domain names in NXDOMAIN packets as input to both
random forests and support vector machines. After training, FANCI can detect malicious traffic in
real-time by analyzing DNS packets as they arrive to the network. 

A central challenge in machine learning approaches for malicious DNS traffic detection is
feature selection. To tackle this issue, deep learning, which can take as input the raw domain
names, has been proposed as an efficient
alternative.  \cite{woodbridge2016predicting,mac2017dga,li2019machine,
tran2018lstm,drichel2020analyzing,sidi2020helix, vinayakumar2020visualized}.


However, since the above works depend on accessing the domain names in the DNS packets, they
cannot be used to detect malicious DoH traffic. The reason is that the domain names are part of
the DNS packet payload that is encrypted by the DoH protocol. 

\subsection{Classification of Encrypted Traffic}
Researchers have also studied classification of encrypted network traffic 
\cite{yang2019bayesian, houser2019investigation,aceto2019mobile, zhang2019deep}.
This research mainly focuses on classifying packet flows according to their application using
deep learning approaches.  However, since they only attempt to classify traffic according to
its application without looking for malicious traffic, these works would simply classify
malicious DoH traffic as an additional DoH packet flow. Nonetheless, the success of these works
in classifying encrypted flows motivate the use of deep learning to detect malicious DoH
traffic.  

\subsection{Detection of Malicious Traffic using an Autoencoder}
In addition to classification of network traffic, 
deep learning has been successfully used to detect malicious traffic in various network environments. 
Meidan et al. \cite{meidan2018n} and Hwang et al. \cite{hwang2019detecting} train autoencoders
that can recreate the normal network traffic of IoT devices. The autoencoders can
detect when the device traffic deviates from the normal one.  
Dargenio et al. \cite{dargenio2018exploring} use autoencoders to identify the most relevant
features for malicious traffic detection.  
Kim et al. \cite{kim2020botnet}  find infected devices with variational autoencoders that
analyze aggregate packet statistics in time windows. 
Zavrak and {\.I}skefiyeli
\cite{zavrak2020anomaly} propose both an autoencoder and a variational autoencoder to detect various
types of malicious network flows using plaintext traffic datasets. 

Our work is the first one to design an autoencoder to detect not only malicious
DoH traffic but any type of malicious DNS traffic. Specifically, 
the autoencoders proposed by the previous works 
\cite{meidan2018n,hwang2019detecting,dargenio2018exploring,kim2020botnet,zavrak2020anomaly}
are trained using datasets that lack both benign and malicious DoH traffic. 
Thus existing autoencoders are not well-suited to detect malicious DoH traffic. 


\subsection{Classification of Malicious DoH Traffic}
There a few works that propose methods to classify malicious DoH traffic.
MontazeriShatoori et al.
\cite{montazerishatoori2020detection} evaluate both machine learning and deep learning
algorithms to classify DoH tunneling data. They propose a
two-step approach. In the first step, their detector separates DoH traffic from the overall
HTTPS traffic by analyzing a set of hand-crafted features from the HTTPS traffic.  The second
step classifies the DoH traffic into malicious and normal.
Singh et al.
\cite{singh2020detecting} and Zhan et al. \cite{zhan2022detecting} employ several
classification models, including naive Bayes, logistic regression, and random forest to detect
DoH tunnels for data exfiltration.  
Kwan et al. \cite{kwan2021exploring} explore how to design DoH tunnels that are difficult to
detect in a censored network. They propose a threshold detection mechanism over a set of
hand-crafted features obtained from the overall HTTPS traffic in a network.
Alenezu et al. \cite{alenezi2021classifying} use recurrent neural networks to classify
different types of malicious DoH traffic used for data exfiltration. 
Vekshin et al. \cite{vekshin2020doh} propose a machine learning approach to classify benign DoH
traffic from other types of HTTPS traffic.

Several enhancements have been proposed to the above malicious DoH traffic classification approaches.
Zebin et al. \cite{Zebin2022AnEA} propose an explainable random forest approach that finds the most
important features for classification. 
Al-Haija et al. \cite{AlHaija2023ALD} design a two-stage approach with minimum number of features. 
Moure-Garrido et al. \cite{MoureGarrido2022DetectingMU} use statistical analysis to identify the
features that have the largest influence. 
Wang et al. \cite{Wang2022FFMRAD} use convolutional neural networks to classify traffic using fused
features as input. 

Existing works on detecting malicious DoH traffic use supervised learning, i.e., they
classify traffic into malicious and benign.
This approach suffers from the following shortcomings. First, the detection performance of this
approach heavily depends on the ability of the defender to collect and maintain an up-to-date
database 
of malicious traffic samples, which usually requires a large amount of resources to collect.
Second, when the available malicious samples are out of date, supervised learning is unable to
accurately classify traffic generated by new malware, i.e., zero-day attacks.
The main reason is that supervised models assign malicious traffic from a
previously unseen class to one of the traffic classes present in the training dataset, possibly
the normal traffic class, which leads to an unpredictable and inaccurate classification
\cite{laskov2005learning, zanero2004unsupervised,usama2019unsupervised,liu2022msca}.

To the best of our knowledge, this is the first work that can detect multiple types of
malicious DoH traffic, even those that are unknown during the training phase.

\section{Background}\label{sec:background}
In this section, we describe the DNS and DoH protocols, and highlight their differences. 
\subsection{DNS}
\begin{figure}[t]
    \centering
    \includegraphics[width=\columnwidth]{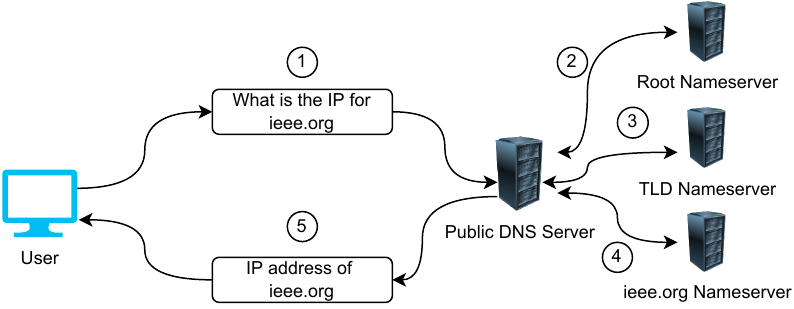}
    \caption{The DNS protocol.}
    \label{fig:normalDNS}
\end{figure}

DNS allows users to find the IP address of servers using human-readable domain names. 
To find the IP address for a given domain name, the DNS protocol
first queries the local device's cache. If the requested IP address is missing from the local
cache, it sends a query to the local DNS server. A local DNS server is usually provided by
the local network or the ISP, but users can choose to use a public DNS server on the Internet.
The DNS server then returns the IP address if it exists in its local cache. Otherwise, it
recursively queries nameservers for each level in the domain. For example, consider
\texttt{www.ieee.org}. The local DNS server will get the IP address of the nameserver in charge
of the top-level domain \texttt{.org} from a known root nameserver. The \texttt{.org}
nameserver will provide the IP address of the nameserver in charge of \texttt{.ieee},
and so on. Eventually, the name server controlling the sub-domain \texttt{www.ieee.org} returns
the requested  IP address to the local DNS server, which then sends the response to the
original host. We show
the DNS query process with a public DNS server in Fig. \ref{fig:normalDNS}.

Since DNS uses UDP (or less commonly TCP) as its transport layer protocols, both queries and
responses are sent in plaintext. This allows network administrators and ISPs to apply the
various approaches described in Section \ref{sec:relatedWorks} that use DNS packet contents for
malicious traffic detection. Unfortunately, the ability to access plaintext DNS packets
compromises the privacy of users because it reveals the websites that they visit as well as
long term Internet usage patterns. 

\subsection{DNS-over-HTTPs}
Aiming to protect user privacy, DoH uses HTTPS to encrypt DNS traffic \cite{dohRFC}. In
contrast to DNS, where the hosts send unencrypted packets to the local DNS server over UDP, DoH
hosts first establish a TLS connection with a public DoH server on the Internet. The IP address
of the DoH server can be obtained from a hard-coded file at the host, or by performing a
traditional unencrypted DNS query. After the HTTPS
connection is established with the DoH server, the host encodes its DNS request into either an
HTTP \texttt{GET} or \texttt{POST} request, and transmits it over the TLS connection
to the DoH server. After the DoH server resolves the query using the traditional DNS protocol,
it replies to the host with an HTTP response message where the body contains the IP address
of the domain, or any of the other DNS protocol messages. The HTTP reply from the DoH server is
also sent over the TLS connection. 

Although DoH protects the privacy of the users, it prevents
intrusion detection systems from accessing plaintext domain names and other DNS packet data
that they need to operate. 

\section{Threat Model} \label{sec:threatModel}
In this section, we first describe in details our assumptions about how malware uses DoH, 
and then formulate the problem of detecting malicious DoH traffic.  

\subsection{Finding Command and Control Servers with Domain Generation Algorithms}
Malware usually communicates with a command and control (C2) server to send exfiltrated data
and receive commands. But before it can do so, it needs to find out the C2 server's IP address.
Unlike legitimate servers that can be easily found by making a conventional DNS query, finding
a C2 requires additional steps due to the precautions that attackers take to avoid being
blocked by ISPs or found by law enforcement. 

Specifically, to evade domain name denylists, 
attackers register their C2 under a domain name generated by a domain generation algorithm
(DGA). The DGA pseudo-randomly generates a list of potential domain names for the C2. The
attacker registers the C2 under one of these domain names. Then, to find the C2, 
infected devices run the DGA to generate the list of potential domain names. Since the infected
devices are unaware of the domain name registered by the attacker, they iteratively send DNS
queries for each potential 
domain name. Eventually, the infected device will query for the registered domain name
and obtain its IP address. If the ISP adds the C2's registered
domain name to the denylist, the attacker can simply register the C2 under different domain name from the list.
The infected devices can follow the same procedure to find the C2 under the new domain name. 
We assume infected devices will use a DGA to find their C2 servers. 

We further assume that the infected devices will attempt to avoid intrusion detection systems
that rely on plaintext DNS queries. Specifically, by employing DoH instead of DNS to find
their C2 server, infected devices can hide their requested domains from the intrusion
detection system. This is a reasonable assumption since security researchers have recently
observed many malware samples using DoH to avoid detection, for example
\cite{godlua,googleAbuse,psixbot,denonia}.
\subsection{Data Exfiltration with DNS}
\begin{figure}[t]
    \centering
    \includegraphics[width=\columnwidth]{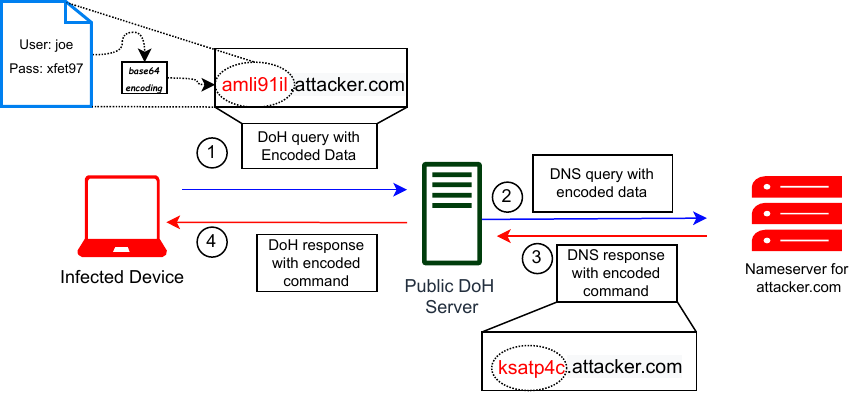}
    \caption{Data exfiltration with DoH tunneling. }
    \label{fig:exfiltrationDoH}
\end{figure}

Besides finding the IP address of their command and control server, malware often uses the DNS
protocol to exfiltrate data in a stealthy manner. The main idea is to use the second-level 
domain of a subdomain owned by the attacker to encode the exfiltrated data. When the infected
devices send a DNS query for this domain name, the DNS query will eventually be forwarded to
the subdomain's nameserver, which is controlled by the attacker.  
Although the second-level domains will not exist, the adversary's nameserver can decode the
data in the second-level domain and reply to the infected device with a DNS packet.  
For example, suppose the subdomain \texttt{attacker.com} is owned by
the adversary and that the data that an infected device wants to exfiltrate is
\texttt{user:joe} and \texttt{password: xfet97}. Then, the infected device would encode the data
and make a DNS request for the second-level domain, e.g., \texttt{amli91il.attacker.org}, where 
\texttt{amli91il} is the encoded data.
The local DNS server eventually sends the query to the attacker-controlled 
nameserver for \texttt{attacker.com}. The nameserver decodes the subdomain \texttt{amli91il} to
obtain the exfiltrated data. We show an example of this process in Fig. \ref{fig:exfiltrationDoH}. 
We assume the infected devices will exfiltrate data using DoH instead of DNS. This is a
reasonable assumption due to recently observed malware that uses DoH for stealthy data exfiltration,
e.g., Qakbot \cite{qakbot}.

\subsection{Problem Formulation}
In this work, we consider the problem of detecting malicious DoH traffic generated by infected
devices  in a local network. Specifically, we define a local network as the set of devices
that are owned by a single person (e.g., home LAN), or organization (e.g., enterprise
network). A network manager sets the list of trusted DoH servers for the network devices, and
obtains the statistical features of network flows by analyzing the packets traversing the
gateway devices. Furthermore, we assume there are only a few approved DoH servers in the network. 
This is justified by the fact that commonly used browsers and operating systems only provide
one to three approved DoH servers by default \cite{mozillaServers,microsoftServers}.

Since legitimate devices only use the approved DoH servers, it is easy for the  network
administrator to differentiate between DoH traffic from other HTTPS traffic, i.e., web
browsing, based on the destination IP address of outgoing packets.  Hence, we focus on
designing a detection approach that uses the encrypted DoH packets as observed by a local
network's gateway device.


Moreover, we assume the malware uses one of the approved
public DoH servers to find its C2 server and to exfiltrate data. 
Consequently, the infected device cannot be easily detected by inspecting the destination IP
address or port number of its outgoing packets since legitimate users will be using the same
server and port number. Note that malware that uses a non-approved DoH server looses the
advantage of communicating with an approved IP address by the network administrator, and thus its
packets can be easily detected and blocked based on the destination IP address.

%

\section{An Autoencoder Architecture for Detection of Malicious DoH Traffic}\label{sec:autoencoder}
In this section, we explain our malicious DoH traffic detection approach and the architecture
of the autoencoder in details. 

\subsection{Overall Approach}
The main objective of the autoencoder is to recreate statistical features from observed
legitimate DoH traffic with high accuracy. If the input to the autoencoder is a legitimate DoH
traffic flow, the reconstruction error will be small. Otherwise, the reconstruction error will
be large.
Thus, we use the magnitude of the reconstruction errors to determine if the observed DoH traffic
is malicious. 

Moreover, 
since we assume that there are only a few DoH servers approved by the local networks, it is
feasible to train an autoencoder for each allowed server. Therefore, to detect malicious
traffic, we filter the DoH traffic by DoH server and then use the corresponding autoencoder to
evaluate it.  

\subsection{Autoencoder Architecture}
\begin{figure}[t]
    \centering
    \includegraphics[width=0.8\columnwidth]{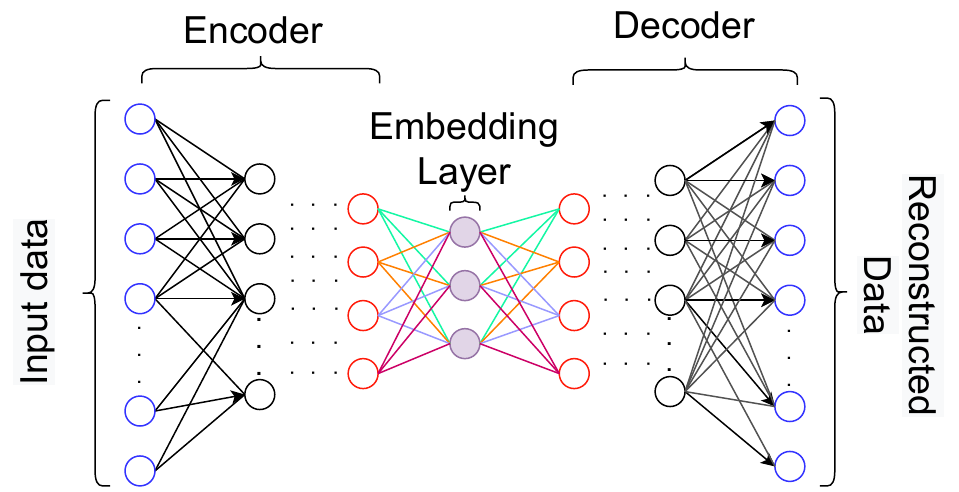}
    \caption{An autoencoder architecture. }
    \label{fig:autoencoder}
\end{figure}

Our autoencoder is formed by two main parts: 1) an encoder neural network that takes as
input the feature vector and outputs a low-dimensional vector, called the embedding; and 2) a
decoder neural network that takes the embedding as input and outputs a reconstruction of the
original feature vector. We illustrate the architecture of the autoencoder in 
Fig. \ref{fig:autoencoder}.

Formally, let $\mathbf{x}\in\mathbb{R}^n$ be the original statistical feature vector obtained
from the DoH traffic flow, $\mathbf{e}\in\mathbb{R}^p$ be the embedding, and
$\mathbf{\hat{x}}\in\mathbb{R}^n$ be the reconstructed statistical feature vector. Then, the
encoder function 
is given by $E(\mathbf{x})=\mathbf{e}$ and the decoder function by
$D(\mathbf{e})=\mathbf{\hat{x}}$.  The complete autoencoder network function
$A:\mathbb{R}^n\rightarrow\mathbb{R}^n$ is defined as $A(\mathbf{x}) = D(E(\mathbf{x}))$.

To train the autoencoder for malicious DoH traffic detection, we train autoencoder $A$ as a
single neural network that aims to output a vector $\mathbf{\hat{x}}$ that reconstructs the input
$\mathbf{x}$. Consequently, the encoder network $E(\mathbf{x})$ learns to create 
an embedding with dimension $p<<n$ that contains enough information for the decoder $D(\mathbf{e})  $
to output an accurate reconstruction $\mathbf{\hat{x}}$. 


\subsection{The layers of the autoencoder}
The layers of the encoder $E$ and decoder $D$ networks are composed by a series of functions
called neurons. Neurons take as input the output from the previous layer and their output is
used as input by the next one. Neurons process their inputs using an activation function
parametrized by weights and biases.

Specifically, let $\mathbf{h}^{k}_E\in\mathbb{R}^{I^k_E}$ be the output of the $k$th
layer of encoder $E$, where $I^k_E$ is the number of neurons. Let
$\mathbf{W}^{k}_E\in\mathbb{R}^{I^k_E\times I^{k-1}_E}$ and
$\mathbf{b}^{k}_E\in\mathbb{R}^{I^k_E}$ be  matrices denoting the weights and biases of the $k$th
layer, respectively. The total number of layers is denoted by $K$. Then, the
output of the $k$th  layer is given by:
\begin{align}\label{eq:ann}
\begin{aligned}
\mathbf{h}^{k}_E&=f\big(\mathbf{W}_E^{k}\cdot \mathbf{h}^{k-1}_E + \mathbf{b}^{k}_E\big), \ \forall
k=1\dots,K,
\end{aligned}
\end{align}
where $f$ is the activation function. The input layer is given by the vector of statistical 
features for a packet flow, i.e., $\mathbf{h}_E^{0}=\mathbf{x}$.  

The size of the layers in encoder $E$ decreases monotonically from the size of the input
layer $n$ to the size of the embedding layer $p$. Thus, we have that $I_E^{k+1}<I_E^{k}$ for $0\leq k
\leq K-1$.

Decoder $D$ is similarly defined. The main difference is that $D$ has one fewer layer
than $E$ (i.e., no embedding layer), and has layers with increasing sizes.  
In particular, decoder $D$'s input layer size $I^0_D$ is given by the size of the embedding
layer $p$, while its output layer $I^{K-1}_D$ is equal to $n$.  
The hidden layers of $D$ have monotonically increasing sizes, i.e., $I_D^{k+1}>I_D^{k}$ for
$0\geq k\geq K-2$.

The exact size of each of the layers and the total number of layers of the encoder $E$ and
decoder $D$ are design parameters that we explore in Section \ref{sec:results}.


\subsection{Malicious DoH Traffic Detection with an Autoencoder}
We use the autoencoder's reconstruction error to determine if the observed DoH traffic
correspond is normal or malicious. Since the autoencoder is trained with normal DoH traffic, we
expect the reconstruction error for malicious traffic samples to be larger than the one
obtained for normal traffic samples. 

The reconstruction error is calculated as the mean squared error between the input and output
features, i.e.,
\begin{align}
  MSE = \frac{1}{n} (x_i - \hat{x}) ^2 
\end{align}
where $x_i$ and $\hat{x}_i$ are the $i$th element of $\mathbf{x}$ and $\hat{\mathbf{x}}$,
respectively.  If a sample has an $MSE$ larger than a threshold $t$, we determine it is
malicious. Otherwise, we determine it is normal. 

The threshold $t$ can be calculated in two ways depending on the availability of malicious
samples. If there are no available malicious samples during training, then $t$ can be calculated
based on the distribution of MSEs for normal samples, i.e., 
\begin{align}
   t =  \mu + s*\sigma
\end{align}
where $\mu$ is the mean of the MSEs for normal samples obtained during training, $s$ is a
positive integer, and $\sigma$ is the standard deviation of the MSEs.  

If malicious samples are available, then we can test several values for the threshold $t$ and
choose the value that offers the best trade-off between the false positives rate and the true
positive rate. This can be achieved by plotting the receiver operating curve (ROC) and calculating
the area under the curve (AUC). Alternatively, the threshold $t$ can be chosen as the value
that offers the best trade-off between precision and recall. In our experiments, we use the ROC
curve to fully explore the detection capabilities of the autoencoder.  

\section{Datasets}\label{sec:dataset}
Our dataset includes various types of normal and malicious DoH traffic. We use this dataset in
Section \ref{sec:results} to train and evaluate the autoencoder. The rest of this
subsection describes the dataset in details. 

\subsection{Normal DoH Traffic Data}\label{sec:datasetnormal}
We collect normal DoH traffic by visiting popular websites using an automated browser. Specifically, 
we program a browser to visit a list of the 1000 most popular websites
\cite{dataforseo} using Python's Selenium library. The browser visited each website multiple times.
Each time using a different public DoH server from either AdGuard, Google, Quad9, or Cloudflare.  
The browser ran inside a virtual machine and all of its network traffic was collected using 
Wireshark. Using the known IP address of the public DoH servers, we filtered the DoH traffic from
the rest of the browser's traffic. This dataset is publicly available \cite{salinas2023dataset}.

We complement our normal DoH traffic data set with traffic
collected by Vekshin et al. \cite{vekshin2020doh} using a DoH proxy server. Their proxy
server translated DNS requests from users in a small office network into DoH requests. They used 
the public DoH server from Cloudflare. 

We use  Python's \texttt{NFStream} library \cite{nfstream} to group the raw packets into
flows, where each flow corresponds to a single TCP connection. Then, for each flow, we calculate
the  average, median, variance, minimum,  and maximum of the number of incoming and outgoing
packets, packet sizes in bytes, connection duration, and inter-packet delays.  In total, we
calculate 16 features for each flow. All features are scaled between zero and one. We later use
these features as the input $\mathbf{x}$ for the autoencoder.  

\subsection{Malicious DoH Traffic Data}
Our malicious DoH traffic data set is formed by encoding DNS queries from DGAs and
tunneling tools into DoH packets. Specifically, we run
20 real-world DGAs that have been reversed engineered \cite{dgaGithub}.  For each domain
generated by a DGA, we form a DNS query, encode it into a DoH query, and then send it to a
public DoH server. We stop generating DNS queries at a randomly chosen domain to simulate an
infected device finding the correct domain for its C2 server. We use Cloudflare's public DoH server to
generate the DGA traffic. This data set is available publicly available
\cite{salinas2023dataset}. 

To simulate malware that attempts to shape its DoH traffic to avoid detection, we
generate malicious DoH requests with different combinations of transmission frequency and
number of TLS connections. In particular, we generate DoH requests for each DGA with the
following characteristics: 
\begin{itemize}
\item \emph{Multiple Connections (MC)}: For each DoH query, we establish a new TLS
connection to the DoH server, resolve the query, and then close the connection. 
\item \emph{Single Connection (SC)}: We establish a single TLS connection for all DoH queries
generated by a DGA. The queries are sent immediately after receiving the reply to the previous
query. 
\item \emph{Single Connection with Random Wait (SC-RW)}: Operates in the same way as SC, except that
it waits a random amount of time between 0 and 2 seconds after receiving the reply to the
previous query before sending the next one. 
\end{itemize}

We also consider malware that uses DoH for data exfiltration. In particular, we incorporate the
malicious DoH traffic samples from MontazeriShatoori et al.
\cite{montazerishatoori2020detection}. Their CIRA-CIC-DoHBrw-2020 data set provides DoH traffic
generated by various DNS tunneling tools, 
namely \texttt{iodine}, \texttt{dns2tcp}, and \texttt{dnscat2}. These tools establish a TCP
connection with an attacker-controlled nameserver by employing the subdomain part of the
requested domain as their communications channel.

Similarly to normal DoH traffic, we use \texttt{NFStream} to group malicious DoH
packets into flows and calculate their features. 

\subsection{Training and Evaluation Datasets}\label{sec:datasets}
We build datasets for each DoH server in our normal traffic dataset. 
To ensure fair comparisons in Section \ref{sec:results}, the size of the
datasets is set to $2,350$ flows to match the server with fewer flows, which in this case is
the DoH proxy server \cite{vekshin2020doh}.  These datasets were divided into $k=5$ folds for
$k$-fold cross-validation. Each fold has $1880$ training flows and $470$ flows reserved for
testing the reconstruction accuracy of the autoencoder. 

We also form an evaluation dataset for each combination of DoH server and type of malicious DoH
traffic. We later use these datasets to evaluate the detection performance of the autoencoders. 
The datasets are formed by combining the flows reserved for testing in each of the
training folds described above with randomly selected samples from the malicious datasets. Each
testing fold has 30\% malicious flows. In total, we form 30 data sets, one for each pair of DoH
server and malicious traffic type. We summarize our testing datasets in Table
\ref{tb:datasets}.  \begin{table*}[th]
   \caption{\label{tb:datasets} The datasets used for evaluating the autoencoder performance.}
   \centering
    \scriptsize
    \begin{tabular}{@{}cccccccc@{}}

    \cmidrule(l){2-8}
        \footnotesize
                                          &                     & \multicolumn{6}{c}{\textbf{Malicious Traffic Types}}                                                                         \\ \cmidrule(l){2-8} 
        \parbox[t]{2mm}{\multirow{6}{*}{\rotatebox[origin=c]{90}{\textbf{DoH Servers}}}} &                     & \textbf{DGA SC}   & \textbf{DGA SC-RW} & \textbf{DGA MC}    & \textbf{\texttt{dns2tcp}}    & \textbf{\texttt{dnscat2}}    & \textbf{\texttt{iodine}}    \\ \cmidrule(l){2-8} 
                                          & \textbf{AdGuard (AG)}    & AG-DGA SC    & AG-DGA SC-RW  & AG-DGA MC     & AG-\texttt{dns2tcp}     & AG-\texttt{dnscat2}     & AG-\texttt{iodine}     \\ \cmidrule(l){2-8} 
                                          & \textbf{Google (G)}     & G-DGA SC     & G-DGA SC-RW   & G-DGA MC      & G-\texttt{dns2tcp}      & G-\texttt{dnscat2}      & G-\texttt{iodine}      \\ \cmidrule(l){2-8} 
                                          & \textbf{Quad9 (Q9)}      & Q9-DGA SC      & Q9-DGA SC-RW        & Q9-DGA MC       & Q9-\texttt{dns2tcp}       & Q9-\texttt{dnscat2}       & Q9-\texttt{iodine}       \\ \cmidrule(l){2-8} 
                                          & \textbf{Cloudflare (C)} & C-DGA SC & C-DGA SC-RW   & C--DGA MC & C--\texttt{dns2tcp} & C--\texttt{dnscat2} & C--\texttt{iodine} \\ \cmidrule(l){2-8} 
                                          & \textbf{Proxy (P)}      & P-DGA SC      & P-DGA SC-RW        & P-DGA MC       & P-\texttt{dns2tcp}       & P-\texttt{dnscat2}       & P-\texttt{iodine}       \\ \cmidrule(l){2-8} 
    \end{tabular}%
    \end{table*}

\section{Experimental Evaluation}\label{sec:results}
In this section, we first describe our autoencoder implementation and training parameters. We
then evaluate its performance under varying hyperparameter combinations, i.e., number of layers
and number of neurons per layer.  We also compare the autoencoder's performance to that of 
local outlier factor (LOF), one-class support vector machine (SVM),
isolation forests (IF), and variational autoencoders. 

\subsection{Autoencoder Implementation and Training Parameters}
We use the TensorFlow library \cite{tensorflow} to implement the autoencoder described in
Section \ref{sec:autoencoder}, as well as variational autoencoders for comparison. We use the
Scikit-learn \cite{scikit-learn} library to implement the LOF, SVM, and IF models with
default settings.  The autoencoder training
and inference are carried out on a High Performance Computing cluster using one Nvidia V100
Tesla GPU along, 8 CPU cores, and 16 GB of RAM. 

We train our proposed autoencoders with different hyperparameter combinations for each training
dataset described in Section \ref{sec:dataset}.  
The number of hidden layers was varied from 2 to 6. The size of the
embedding layer was varied from 3 to 9 neurons. The number of neurons of the encoder's hidden
layers were varied from 14 to 110. The input layer was kept constant at 16 neurons to match the
number of input features. At the encoder, the pattern was reversed.  

We use the ADAM algorithm to train the autoencoder and use MSE as the
loss function. Since the goal of the autoencoder training is to minimize the difference between
its input and output, we set the sample labels $\mathbf{x}'$ equal to the sample itself, i.e.,
$\mathbf{x}=\mathbf{x}'$.   The batch size is set to $32$ and the learning rate is kept at the
default value of $0.001$. The number of epochs is set to $30$.   To improve the numerical
stability of the training algorithm and improve reconstruction accuracy, we add batch
normalization layers to the autoencoder. All datasets are scaled to the range $[0,1]$.


\subsection{Autoencoder Training Results}
\begin{figure}[] \centering  
\subfloat[][]{
\includegraphics[width=0.49\columnwidth]
{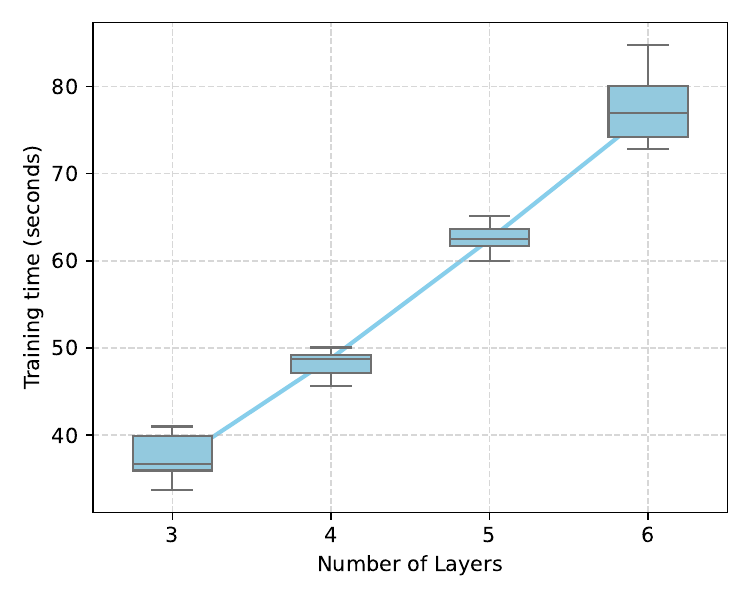 }
\label{fig:training-times}
}
\subfloat[][]{\includegraphics[width=0.49\columnwidth]
{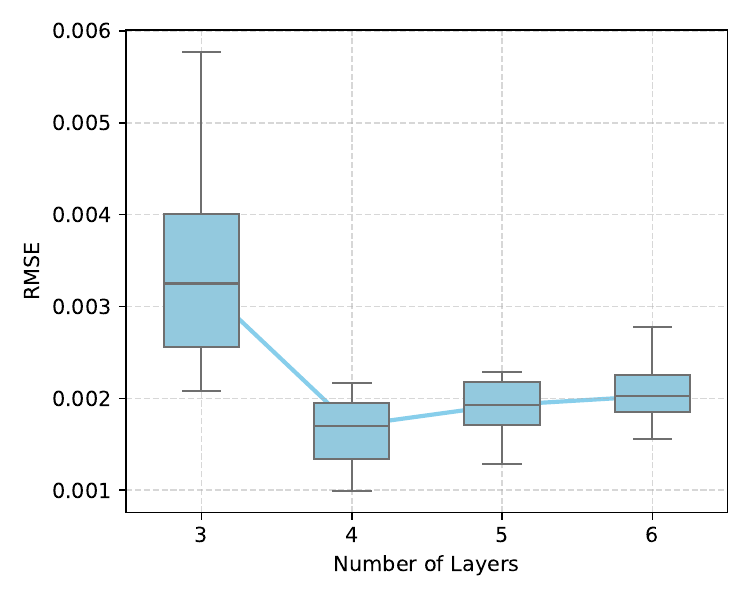}
\label{fig:rmse-training}
}
\caption{Training results: (a) Autoencoder training time for increasing number of hidden layers;
(b) Median RMSE across all architectures and training datasets.   \label{fig:trainingResults}}
\end{figure}

Fig. \ref{fig:training-times} shows the median training time of the autoencoder for an
increasing number of layers across all training datasets. 
We observe that the training time increases with the number of layers. The training
times range from $35$s to $76$s, which is low since the autoencoder would only need to be infrequently
updated with new normal DoH traffic.

In Fig. \ref{fig:rmse-training}, we show the mean reconstruction median squared error (RMSE) for
an increasing number of layers across all training datasets. We see that the median RMSE remains
between $0.2300$ and $0.2305$, while the standard deviation is larger for models with 6 layers. This
RMSE values are low, which shows that all the models are able to accurately
reconstruct the normal DoH traffic from the multiple DoH servers.  

\subsection{Performance Comparison between Autoencoder Architectures}
After training the autoencoders, we evaluate their ability to detect malicious DoH traffic in
terms of their accuracy, area under the curve (AUC), accuracy, recall, precision, and F-1
score.  We evaluate the median performance of the different hyperparameter combinations across 
the 30 normal and malicious DoH traffic datasets described in Table
\ref{tb:datasets}. We summarize the evaluation results for the top-ten autoencoder models and
rank them according to their F1-score in Table \ref{tb:architecturePerformance}. The \textbf{Architecture}
column shows the number of layers in the encoder. The layers at the decoder have the same
number of neurons but in reverse order
\begin{table*}
\footnotesize
\centering
\caption{ Metric scores (\emph{median}             $\pm$ \emph{std}) for the top-ten autoencoder models ordered by F-1 score.}
\label{tb:architecturePerformance}
\begin{tabular}{lllllll}
\toprule
\toprule
\thead{Architecture} & \thead{F1-score} & \thead{Accuracy} & \thead{AUC} & \thead{Precision} & \thead{Recall} \\
\midrule
\midrule
16, 62, 9 & $0.99359 \pm 0.03494$ & $0.99018 \pm 0.04056$ & $0.99564 \pm 0.04657$ & $0.99826 \pm 0.00431$ & $0.98825 \pm 0.05210$ \\
\midrule
16, 26, 17, 9 & $0.99359 \pm 0.01912$ & $0.99018 \pm 0.02718$ & $0.99491 \pm 0.02731$ & $0.99830 \pm 0.00287$ & $0.98936 \pm 0.03484$ \\
\midrule
16, 98, 9 & $0.99343 \pm 0.04314$ & $0.98990 \pm 0.05014$ & $0.99571 \pm 0.05929$ & $0.99787 \pm 0.01436$ & $0.98849 \pm 0.06198$ \\
\midrule
16, 86, 9 & $0.99325 \pm 0.02001$ & $0.98966 \pm 0.02777$ & $0.99542 \pm 0.02812$ & $0.99829 \pm 0.00805$ & $0.98846 \pm 0.03410$ \\
\midrule
16, 62, 35, 9 & $0.99287 \pm 0.01568$ & $0.98909 \pm 0.02279$ & $0.99506 \pm 0.02294$ & $0.99827 \pm 0.00147$ & $0.98738 \pm 0.03002$ \\
\midrule
16, 38, 23, 9 & $0.99283 \pm 0.01647$ & $0.98902 \pm 0.02380$ & $0.99499 \pm 0.02381$ & $0.99806 \pm 0.00296$ & $0.98797 \pm 0.03101$ \\
\midrule
16, 26, 20, 14, 9 & $0.99282 \pm 0.02329$ & $0.98901 \pm 0.03291$ & $0.99555 \pm 0.03117$ & $0.99828 \pm 0.00485$ & $0.98740 \pm 0.04255$ \\
\midrule
16, 74, 38, 3 & $0.99282 \pm 0.01353$ & $0.98902 \pm 0.01962$ & $0.99414 \pm 0.01799$ & $0.99856 \pm 0.00211$ & $0.98700 \pm 0.02589$ \\
\midrule
16, 50, 29, 9 & $0.99266 \pm 0.02422$ & $0.98878 \pm 0.03343$ & $0.99477 \pm 0.03248$ & $0.99809 \pm 0.00203$ & $0.98706 \pm 0.04382$ \\
\midrule
16, 110, 56, 3 & $0.99264 \pm 0.01655$ & $0.98874 \pm 0.02376$ & $0.99384 \pm 0.02121$ & $0.99831 \pm 0.00156$ & $0.98756 \pm 0.03137$ \\
\bottomrule
\bottomrule
\end{tabular}
\end{table*}

We see that all the top-ten autoencoder models achieve F1-scores, accuracy, AUC, precision and
recall greater than $0.9800$, which is high. The architecture
$[16, 62, 9]$ achieves a median F1-score of $0.99359$ with a median training time of
$38.8916$ seconds and a median RMSE of $0.0021$ during testing. Although it has a slightly
higher F1-score variance than the next best performing architecture, its simpler architecture
(i.e., only  one hidden), and lower training time make it a more attractive choice. 
For this reason, we use the architecture $[16, 62, 9]$ to
compare our proposed autoencoder to existing anomaly detection approaches in the next section. 

\subsection{Performance Comparison with other Anomaly Detection Models}
We compare the performance of our proposed autoencoder to other models that
have successfully been used to perform anomaly detection in network traffic problems
\cite{falcao2019quantitative,zavrak2020anomaly}. We provide a brief description of these models
next. 

\subsubsection{Variational Autoencoders (VAE).} VAEs are autoencoders with a probabilistic embedding
layer. Instead of finding an embedding layer with deterministic outputs, it instead finds an
embedding layer that defines the latent probability distribution. The decoder network of a VAE
takes as input samples from the latent probability distribution and outputs the probability
that the input was generated by the latent probability defined by the embedding layer. If the
probability is low, the VAE declares the input as an anomaly. Anomaly detection VAEs are proposed by
An and Cho \cite{An2015VariationalAB}. We implemented VAEs with varying number of layers and neurons
per layer. The results in this section correspond to the architecture $[ 16, 50, 26, 3]$, which is
the best performing one. 

\subsubsection{Isolation Forest (IF).} IF is an unsupervised method that builds isolation trees
where leaves represent data points. IF assumes that there are significantly fewer anomalous
samples in the data set and that these samples have drastically different features from the
normal samples. Hence, anomalous samples will have a much shortest path to the root of the tree
compared to the normal ones  \cite{liu2008isolation}. 

\subsubsection{One-class Support Vector Machine (OC-SVM).} The One-class SVM  is a semi-supervised
method that aims to find a decision boundary on a hyperplane with maximum separation between
the normal samples and the origin when only normal data is available, which is the case in our
problem \cite{amer2013enhancing}. OC-SVM can be considered as a type of supervised learning
that classifies normal network traffic as normal, and classifies all other types of traffic as
anomalous.

\subsubsection{Local Outlier Factor (LOF).} LOF is an unsupervised method that calculates an
isolation factor for each sample in the data set. The isolation factor is calculated as the
distance of the sample to the center of its nearest cluster. By calculating the factor relative
to a cluster, the LOF can identify anomalous samples even when there are multiple clusters of
normal data. \cite{breunig2000lof}.

\begin{figure}[t]
    \centering
    \includegraphics[width=0.8\columnwidth]{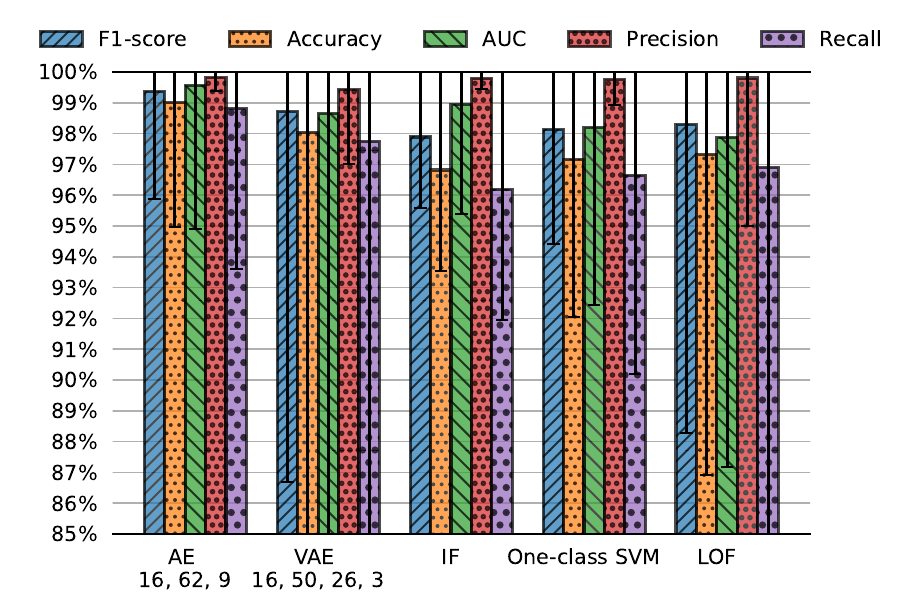}
    \caption{Performance comparison between an autoencoder (AE) and other anomaly detection models. The bar heights are the median values and the error bars are the standard deviations across all benign and malicious DoH traffic datasets.}
    \label{fig:comparisonBar}
\end{figure}

Fig. \ref{fig:comparisonBar} shows the performance metrics for our autoencoder model as well as
for IF, One-class SVM, LOF, and VAE across all normal and malicious DoH traffic datasets.  
Specifically, the bar heights are the median values of the
performance metrics across all datasets. The figure also shows the standard deviation values as
error bars. We see that our proposed autoencoder achieves the highest median F1-score, accuracy,
AUC, and recall. The standard deviation of these metrics is also smallest for the autoencoder.
The median precision is comparable for all the models, with the standard deviation being the
smallest for the autoencoder.

\begin{figure*}[ht]
  \centering
    \subfloat[][AdGuard server]{
    \includegraphics[width=0.65\columnwidth]
    {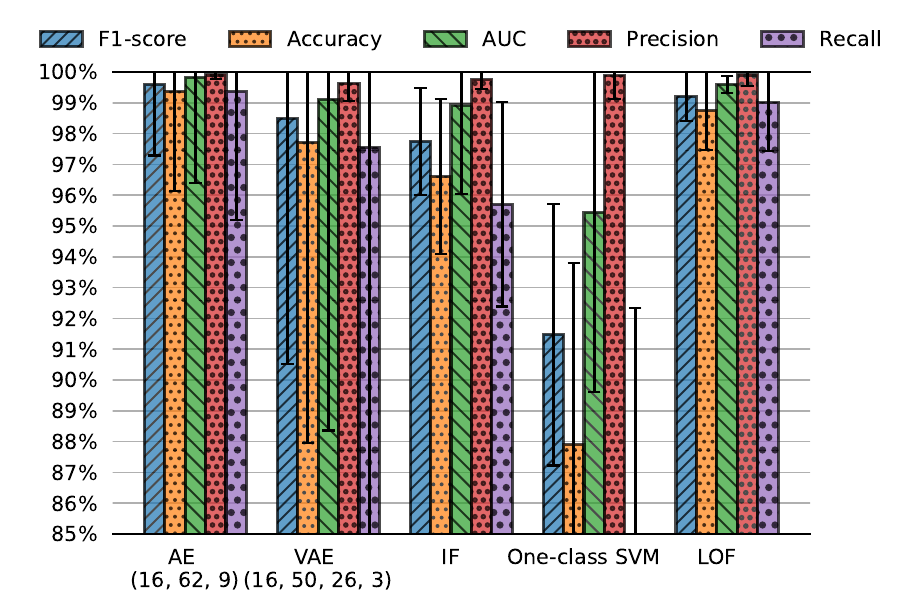
     }}
     \subfloat[][Cloudflare server]{\includegraphics[width=0.65\columnwidth]
     {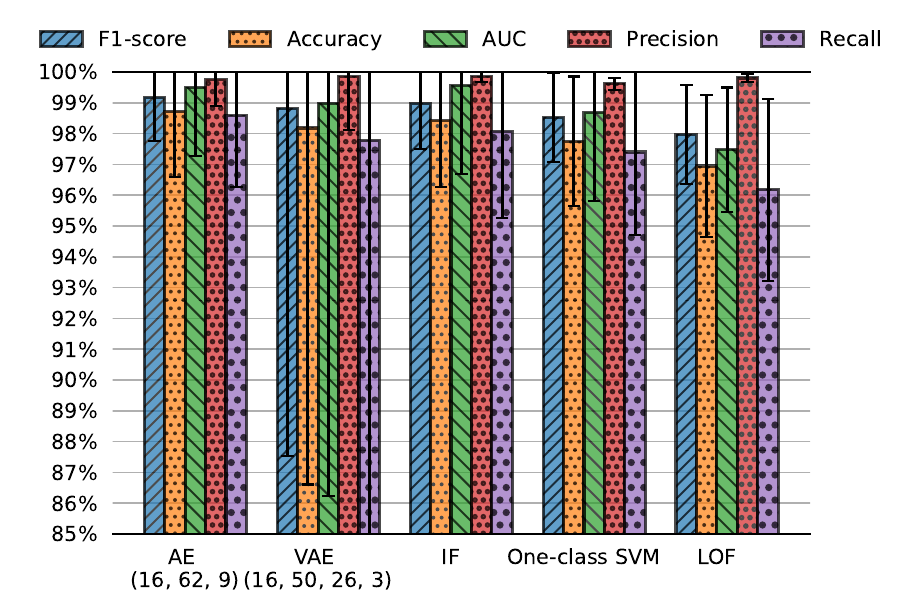
  }}
 \subfloat[][Google server]{\includegraphics[width=0.65\columnwidth]
 {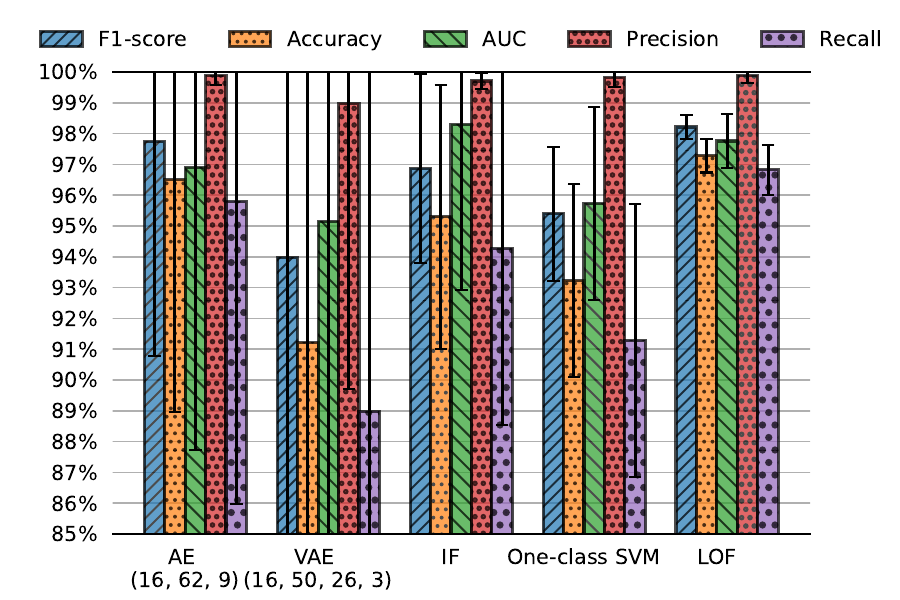
  }}

  \subfloat[][Proxy]{\includegraphics[width=0.65\columnwidth]{
    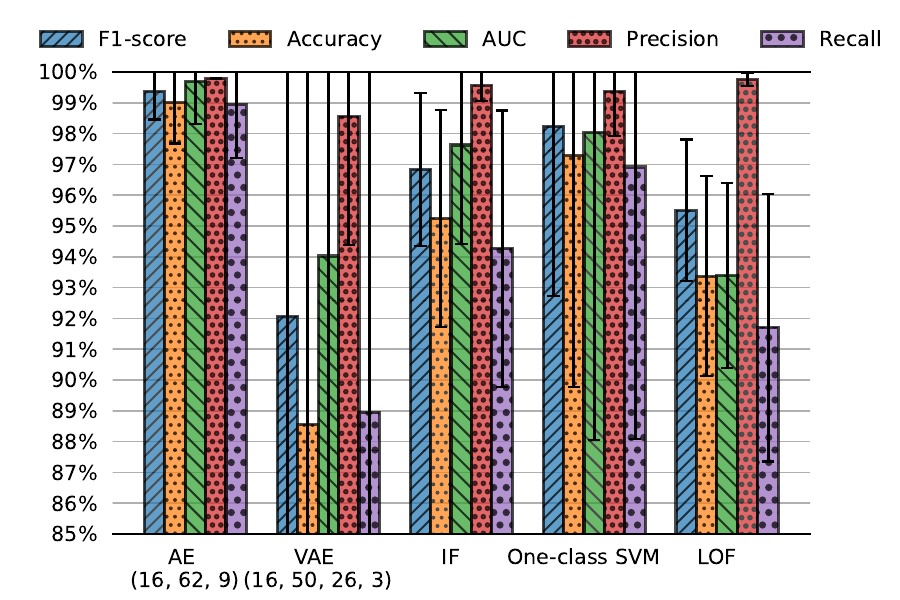
  }}
     \subfloat[][Quad9 server]{\includegraphics[width=0.65\columnwidth]{
    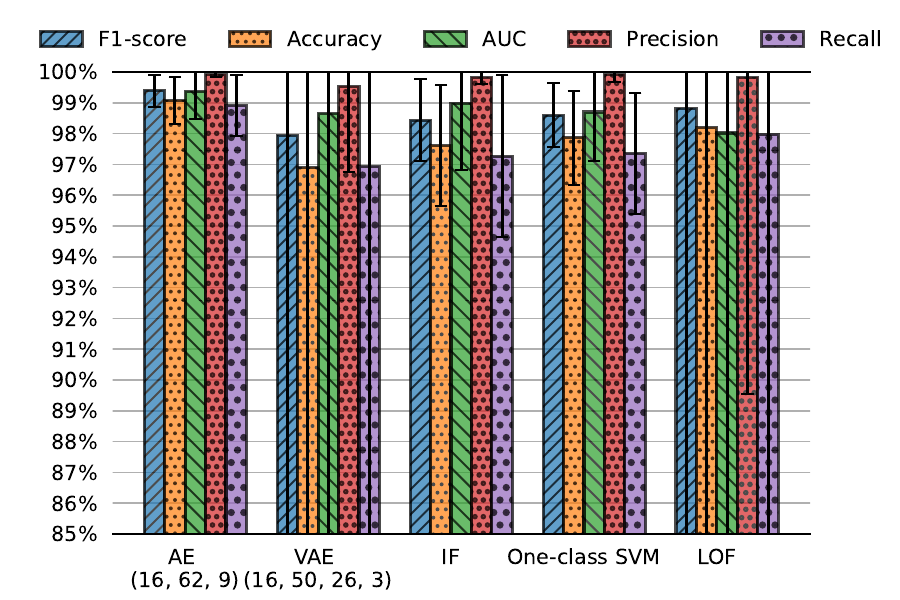
  }}
  \caption{Performance comparison between the autoencoder (AE) and other anomaly detection models
  for a given benign DoH traffic dataset.
  The bar heights are the median values and the error bars are the standard deviations across all
  malicious DoH traffic datasets.  \label{fig:comparisonTraining}}
\end{figure*}

Next, we investigate the impact of training the autoencoder with normal traffic  from
different DoH servers. Fig. \ref{fig:comparisonTraining} compares the median detection
performance of the models across multiple malicious DoH traffic datasets for a given normal DoH
server traffic dataset.  Overall, we see that our proposed autoencoder is the least affected model by
the choice of training dataset.  The autoencoder performs better than the other models for all
datasets except for the Google DoH server dataset, where it performs similarly to LOF. In
contrast, we see that the other models suffer from a significantly lower detection performance
under at least one training dataset compared to the rest.  That is, One-class SVM performs poorly
under the AdGuard dataset, LOF under the Cloudflare and Proxy datasets, IF under the Proxy
dataset, and VAE under the Proxy dataset. These results show that the autoencoder is less
sensitive to the choice of training data sets, and thus can be deployed more widely.  

\begin{figure*}[ht]
  \centering
\subfloat[][DGA Single Connection (SC)]{\includegraphics[width=0.65\columnwidth]{
   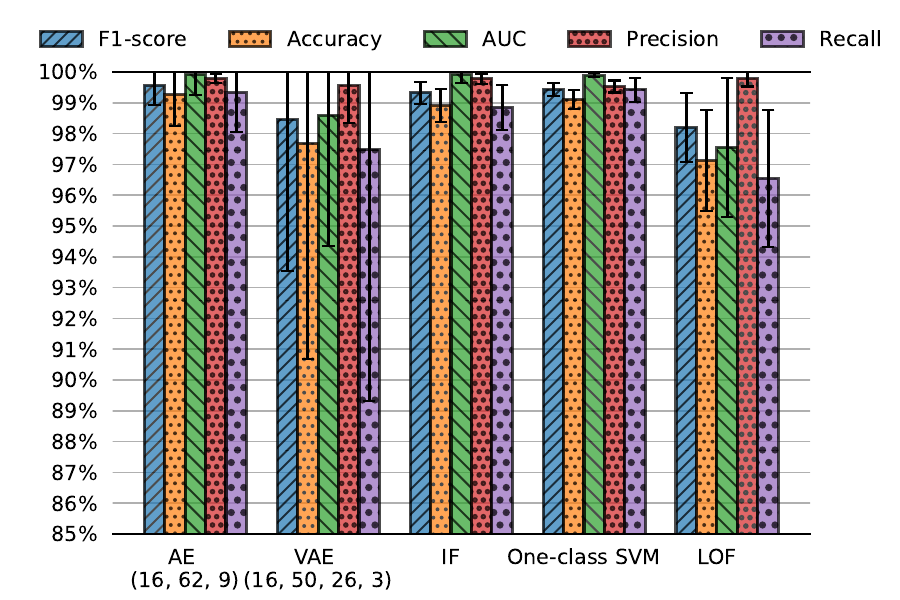
  }}
\subfloat[][DGA Single Connection with Random Wait (SC-RW)]{\includegraphics[width=0.65\columnwidth]{
    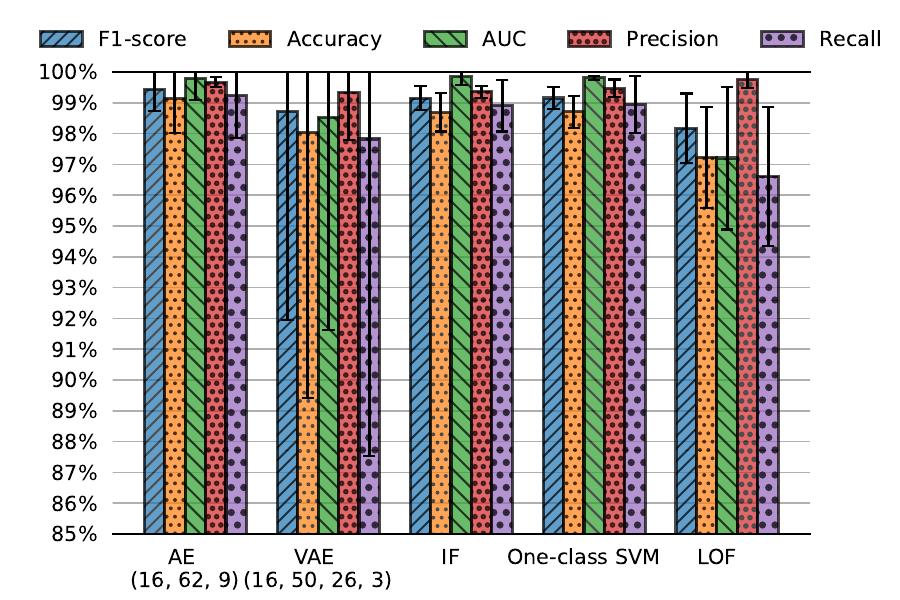
  }}
\subfloat[][DGA Multiple Connections (MC)]{
  \includegraphics[width=0.65\columnwidth]{
  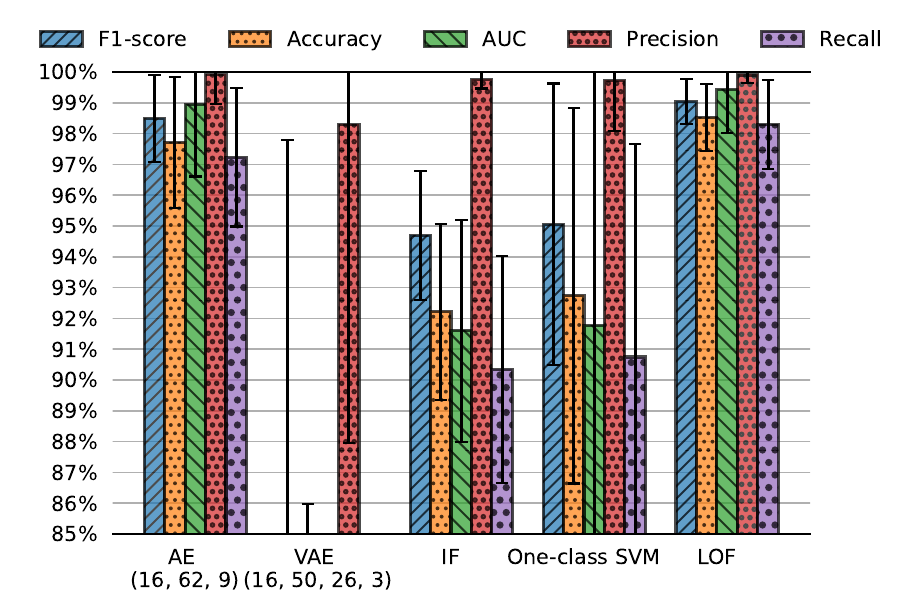
  }}

  \subfloat[][\texttt{dns2tcp}]{\includegraphics[width=0.65\columnwidth]{
    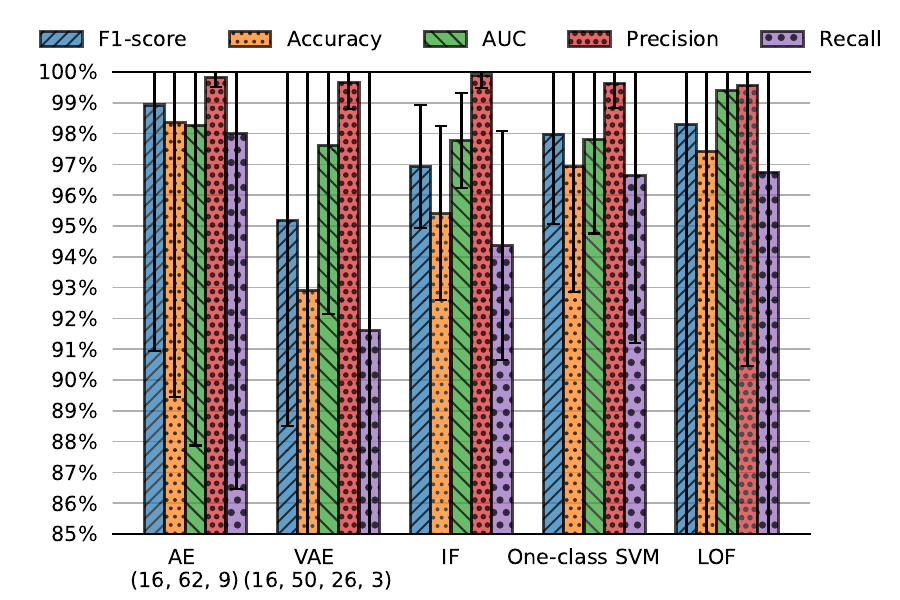
  }}
  \subfloat[][\texttt{dnscat2}]{\includegraphics[width=0.65\columnwidth]{
    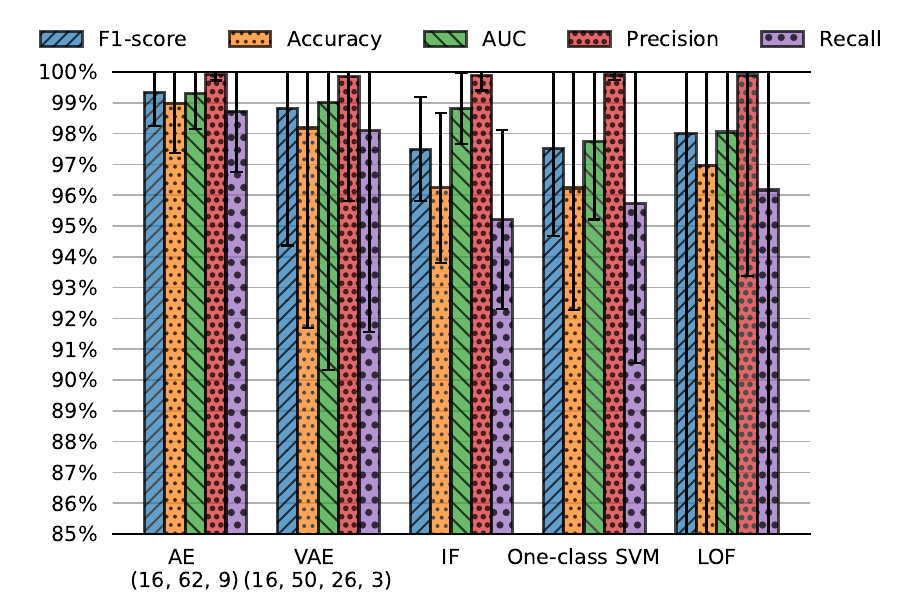
  }}
 \subfloat[][\texttt{iodine}]{\includegraphics[width=0.65\columnwidth]{
    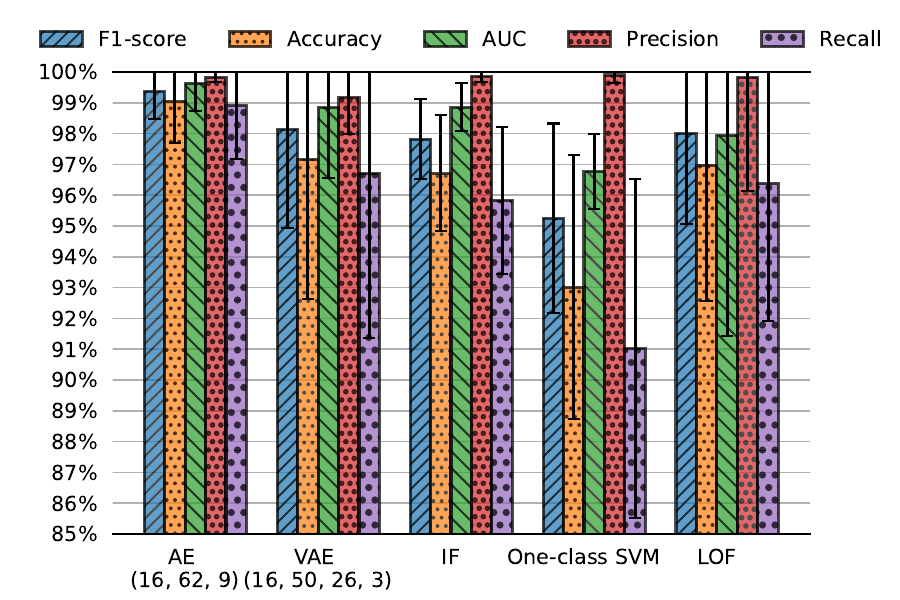
  }}
  \caption{Performance comparison between an autoencoder (AE) and other anomaly detection models.
  The bar heights are the median values and the error bars are the standard deviations across all
  training and testing files. \label{fig:comparisonTraining2}}
\end{figure*}

We now investigate the performance of the models in detecting the different types of malicious
traffic. Fig.  \ref{fig:comparisonTraining2} shows the median performance metrics for a given
malicious traffic type across the normal DoH traffic datasets. We see that the autoencoder
achieves a performance equal or higher than the other
detection models for all malicious DoH traffic types, except for DGA multiple connections where LOF
has a better performance. The autoencoder achieves its highest performance when detecting DGA
SC traffic, and its lowest one under DGA MC. The reason is that DGA
SC traffic contains a high number of requests from each DGA which provides enough
information to determine it is malicious. On the other hand, flows in the DGA MC traffic 
only contain information about a single DoH query, and thus are harder to detect. 

\begin{figure*}
  \centering
  \subfloat[][Autoencoder (16, 62, 9)]{\includegraphics[width=0.5\columnwidth]{
  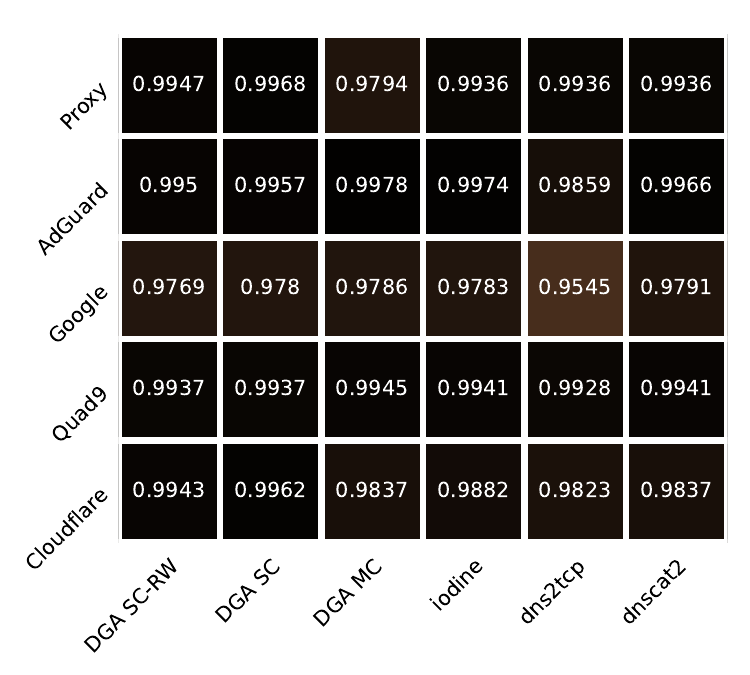 }\label{fig:heatmap-ae} } 
  \subfloat[][LOF]{
    \includegraphics[width=0.5\columnwidth]{
      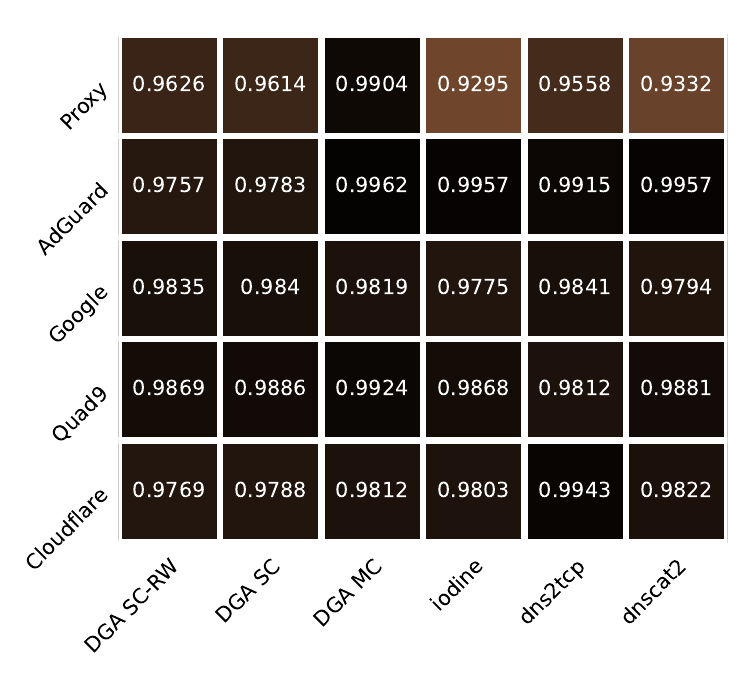
     }\label{fig:heatmap-lof}}
  \subfloat[][IF]{\includegraphics[width=0.5\columnwidth]{
    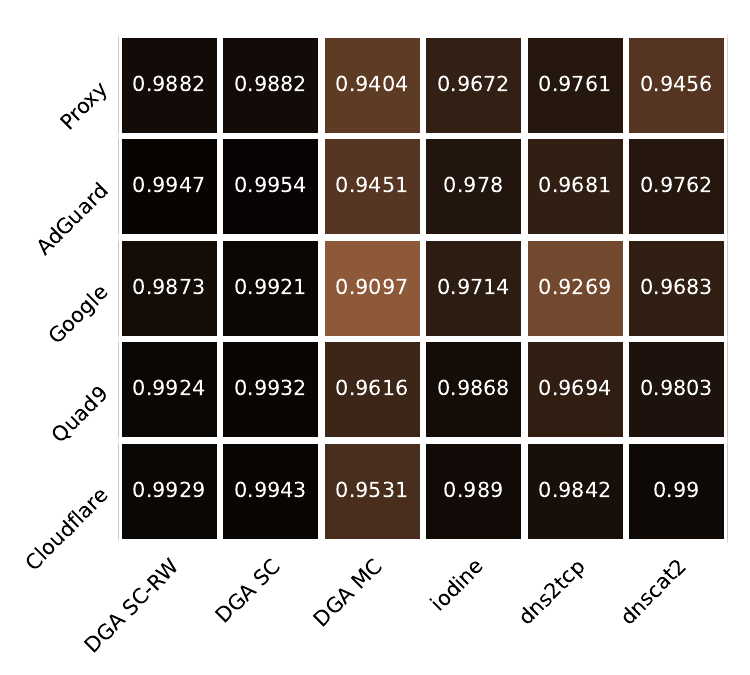
  }\label{fig:heatmap-if}}
  \subfloat[][One-class SVM]{\includegraphics[width=0.5\columnwidth]{
    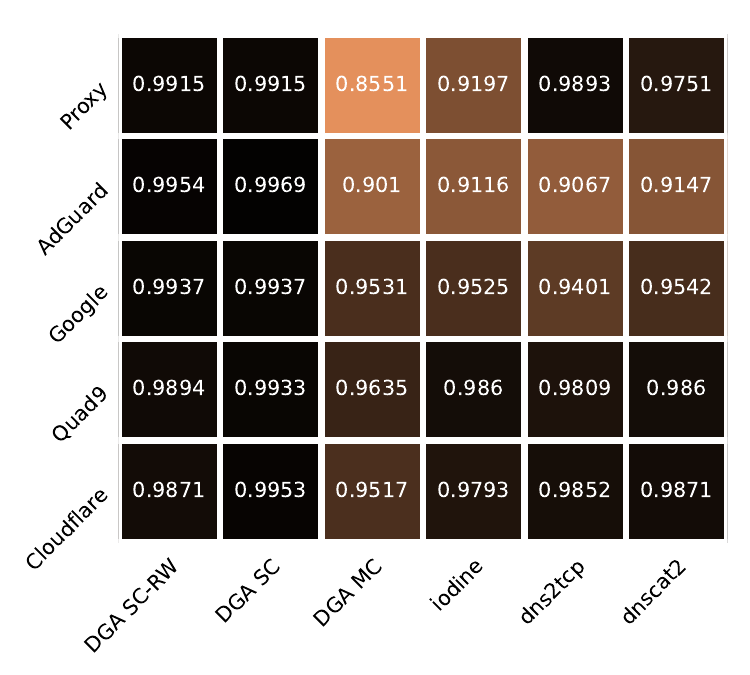
  }\label{fig:heatmap-svm}}
  \caption{F1-scores achieved when detecting malicious DoH traffic under different normal and
   malicious data sets.\label{fig:heatmap-single-datasets}}
\end{figure*}

Finally, we take a deeper look into the autoencoder's performance by considering one evaluation
dataset at a time.  Fig.  \ref{fig:heatmap-ae} shows a heatmap of the F-1 scores achieved by the
autoencoder when trained
with the normal DoH traffic dataset of the corresponding row and tested using normal samples 
and the malicious traffic indicated in the corresponding column. Darker colors
indicate a higher F-1 score, and thus a better overall detection performance for the traffic
combination.  Fig. \ref{fig:heatmap-lof}, Fig. \ref{fig:heatmap-if}, and Fig.
\ref{fig:heatmap-svm} show the F1-scores for LOF, IF, and One-class SVM, respectively. 

Overall, the heatmaps confirm our observation that the autoencoder generally outperforms the
other models, except for the DGA MC.  Although the median F1-score of the autoencoder under the
DGA MC dataset is lower than the score for LOF, we see from Figs. \ref{fig:heatmap-ae} and
\ref{fig:heatmap-lof} that the performance of the models is very close, except for the Proxy
server dataset, where the autoencoder performs worse. When the autoencoder  trained with Google
DoH server traffic attempts to detect \texttt{dns2tcp} traffic, it also achieves a relatively
low F1-score. The VAE performs worse than all the other anomaly detection algorithms, thus it 
is omitted for brevity. 

Note that to maintain a fair comparison, we leave out supervised learning approaches, e.g.,
traditional SVM, classification neural networks, random forest, etc., from our evaluation. The
reason is that supervised learning approaches assume that malicious traffic observed during testing
is generated by a class of malicious traffic available in the training dataset, which is not
the case in our considered scenario. Moreover, we know from the literature
\cite{laskov2005learning,zanero2004unsupervised,usama2019unsupervised,liu2022msca},
that supervised learning performs poorly when classifying traffic from previously unseen classes. 

\section{Conclusions}\label{sec:conclusions}
The DNS protocol compromises users' privacy by revealing their web browsing history to devices
that handle their packets. To address this issue, popular web browsers have adopted
DNS-over-HTTPS (DoH), an encrypted version of the DNS protocol that protect users' privacy.  
Unfortunately, the adoption of DoH prevents existing intrusion detection approaches from using
the plaintext domain names in DNS query packets to detect malicious traffic.  In this paper, we
have designed an autoencoder that detects malicious traffic by only observing the encrypted DoH
traffic. Since the autoencoder was trained to recognize normal DoH traffic, it can recognize
previously unseen malicious traffic.  We run extensive experiments with real-world and
realistic simulated data to evaluate the performance of the autoencoder, and compare it to
other anomaly detection algorithms, i.e., local outlier factor (LOF), one-class support vector
machine (SVM), isolation forests (IF), and variational autoencoders.  We find that the
autoencoder achieves the highest malicious DoH traffic detection performance, with a median F-1
score of 0.99359 across all
training and evaluation datasets. Future work includes using the output of the embedding layer as a
low dimensional representation of the DoH traffic to find anomaly detection models with lower
computing requirements for both training and inference. 

%
%
\bibliographystyle{IEEEtran}
\bibliography{references.bib}


\end{document}